\documentclass[aps,twocolumn,amsmath,amssymb,prl]{revtex4-1}
\usepackage{amssymb}
\usepackage{amsmath}
\usepackage{graphicx}
\usepackage[table]{xcolor}
\usepackage{epsfig}
\usepackage[utf8]{inputenc}
\usepackage{dcolumn}
\usepackage{bm}
\usepackage{array}
\usepackage{multirow} 
\begin{document}

\title{A microscopic model for hydrated biological tissues}

\author{E. T. Sato$^{1}$, A. R. Rocha$^{2}$, L. F. C. S. Carvalho$^{3}$, J. D. Almeida$^{4}$, H. Martinho$^{1}$}

\email{herculano.martinho@ufabc.edu.br}

\affiliation{$^{1}$Centro de Ciências Naturais e Humanas, Universidade Federal do ABC,  Santo André-SP, Brazil\\
$^{2}$Instituto de Física Teórica, Universidade Estadual Paulista (UNESP), São Paulo, Brazil\\
$^{3}$FOCAS Institute, Dublin Institute of Technology, Camben Row, Dublin 8, Ireland\\
$^{4}$Depto. de Biociências e Diagnóstico Bucal, Instituto de Ciência e Tecnologia, Campus São José dos Campos, Universidade Estadual Paulista (UNESP), São José dos Campos-SP, Brazil}

\begin{abstract}

The present work presents a density-functional microscopic model of soft biological tissue. The  model was based on a prototype molecular structure from experimentally resolved collagen peptide residues and water clusters and has the objective to capture some well-known experimental features of soft tissues.  It was obtained the optimized geometry, binding and coupling energies and dipole moments. The  results concerning the stability of the confined water clusters, the water-water and water-collagen interactions within the CLBM framework were successfully correlated to some important trends observed experimentally in inflammatory tissues.

\end{abstract}

\maketitle

The role of water in biological systems is still in many case an enigma. Water is an integral part of many biomolecules. In particular, proteins have much of their properties governed by interactions with water\cite{chaplin}. Leikin \textit{et al.}\cite{leikin1997raman} have shown that hydration force effects resulting from energetic cost of water rearrangement near the collagen surface display distinguished features in the Raman spectra of this molecule. Moreover, in a previous study of our group\cite{e2011diagnosis}, it was shown that water behaves differently in healthy and pathological tissues. By analysis of the vibrational modes compared to the results presented in the literature\cite{movasaghi2007raman,cybulski2007calculations}, it was observed that the main changes of this inflammatory process were related to collagen and confined water in the high-wave number region. Furthermore, the intracellular hydration can be of great relevance in cancerous processes, since cancer cells contain more free water than normal cells, and the degree of malignancy increases with the degree of cellular hydration\cite{mcintyre2006cell,chaplin2006we}. Hydropic degeneration (excess of water absoption by the epithelial tissue) and biochemical redox reactions are two important characteristics of the inflammatory process. The prevalence of typical water clusters in inflammatory tissues is a known event reported in the literature \cite{martinholivro}. Cibulsky and Sadlej\cite{cybulski2007calculations} observed that Raman spectra in the OH-stretching vibrations region are related to local structures and interactions of the hydrogen-bond networks, so each cluster has a characteristic Raman spectrum. de Carvalho \textit{et al.}\cite{e2011diagnosis} pointed out that this could be origin of the discriminative power of the high wavenumber region closely related to the hydropic degeneration process in inflammatory fibrous hyperplasia.  Gniadecka \textit{et al.}\cite{gniadecka1997diagnosis} concluded that an increased amount of the non-macromolecule bound, tetrahedral water was found in photoaged and malignant tumours of the skin. These experimental results indicate the relevance of understanding the role of electrons, protons and hydration water in tissues.

\begin{figure}[h!]
\includegraphics[width=7.0cm]{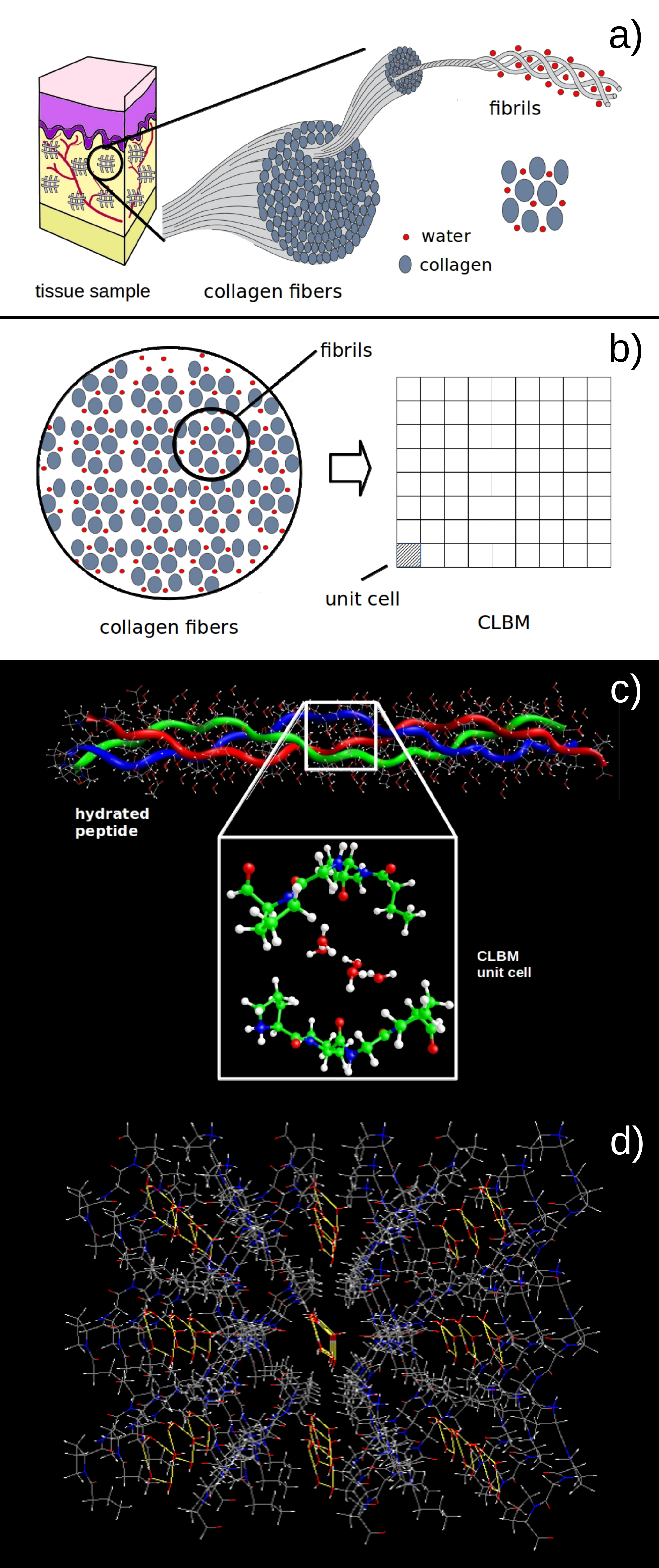}
\caption{a) Structural scheme of a conective tissue showing the collagen fibers and the fibrils. These are constituted by water and collagen molecules b) The construction of the  CLBM unit cell compared the original collagen fibers structure. c) The CLBM unit cell chosen from a experimentally resolved hydrated peptide structure. d) CLBM enclosing 4 water molecules clusters.  The hydrogen bondings inside the water clusters are indicated by yellow. Hydrogen, carbon, nitrogen, and oxygen atoms were in white, green, blue, and red, respectively.}\label{clbm}
\end{figure}

Computational simulations are widely used to study and make predictions concerning a broad variety of systems ranging from pharmacology to engineering fields\cite{comp_pharma,comp_matsci,comp_mech}. Atomistic models based on quantum mechanics calculations have the greatest materials properties predictive capability. However, due to their inherent complexity detailed atomistic models for biological tissues is absent in the literature. Such model would be useful to understand physical/biochemical properties of tissues. For example, little is known about the underlying mechanisms of cell migration in wound healing specially the  modulating role of the mechanical tension at microenviromental scale\cite{lu2013asymmetric}. Usually finite-element classical approach to model soft tissues has been applied to simulate mechanical characteristics of soft tissues like skin\cite{Groves2013167}. However, many physical properties have not been satisfactorily simulated by these conventional models\citep{MC1,MC2}. Monte Carlo methods has been generally considered as the gold standard of modeling light propagation in heterogeneous tissues and used to validate the results obtained by other models. But the main drawbacks of this method are the extensive computational burden \cite{MC1}, absence of realistic phase function and elastic light scattering models\citep{MC2}.

Computer simulations however, the dynamics of electrons, protons and hydration water using quantum mechanical approach as, to best of our knowledge, have only been studied for isolated or solvated molecules. Of all proteins present in our body, collagen is the most abundant and for this reason has been exhaustively studied. Though, theoretical models exist, they are mostly treated classically \cite{Mogilner2002209,DeSimone2008121,terimd,C0SM01192D}, although there are also hybrid models \cite{suarezqmmm}.

In this work a microscopic model based on Density Functional Theory (DFT) for soft biological tissue is presented. This model, named collagen like bulk model (CLBM) retains some fundamental atomic connectivity based on the collagen fibrils constitution. The stability of some proposed variations of the model was studied and compared to selected experimental results.  

Figure \ref{clbm}a) shows a general scheme of a real connective tissue. The different collagen types form larger fibrillar bundles. Collagen fibrils are semi-crystalline aggregates of collagen molecules, which are in form bundles of fibrils.  Water and large collagen molecules ($> 1,000$ amino acids) are the main constituent of the fibrils. The model purpose is to simulate a tissue; the set of collagen fibers and fibrils in the presence of water molecules between them. Figure \ref{clbm} b) helps to show the dimension scale where our model was built. We associate the  collagen fibers set to a reticulated set of unit cells whose internal constitution will be chosen. In our approximation the complexity of the tissue will be replaced by this periodic set using periodic boundary conditions. The CLBM unit cell was built starting with an experimentally resolved hydrated collagen type I structure obtained from the Protein Data Bank\cite{bella1995hydration}.  A specific peptide sequence of this structure was cut retaining proline, glycine and hydroxyproline amino acids which enclose some water molecules already mentioned (Fig. \ref{clbm}c) . The criteria used to choose this specific peptide sequence was based on (i) the presence of water molecules confined in the original structure; (ii) presence of residues of amino acids; (iii) viable set of amino acids with respect to time of calculation in DFT (typically $\sim 100$ atoms). Here, we are not only simulating the collagen molecule, but the set of tissue fibers.  

\begin{figure}[h!]
\includegraphics[width=8.2cm]{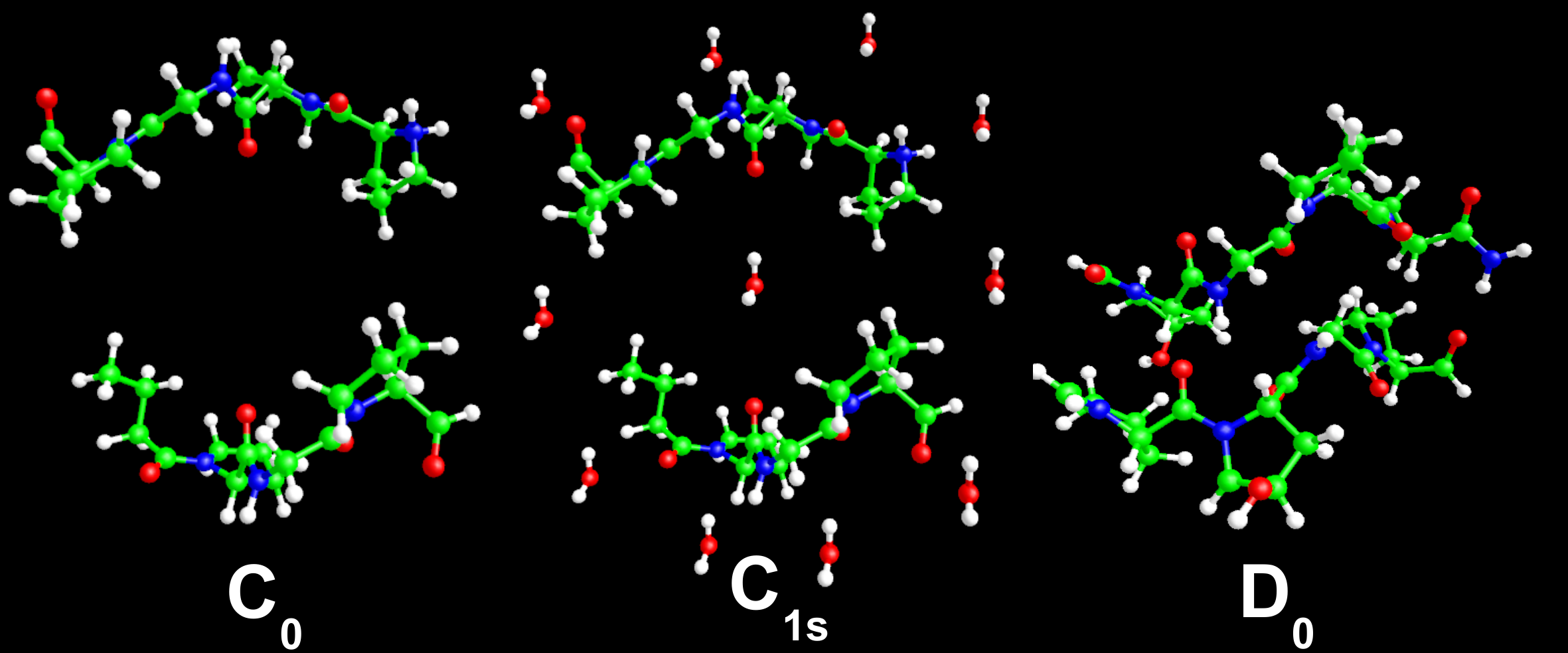}
\caption{$C_{0}$, $C_{1s}$, and $D_{0}$ structures.}\label{c1sd0}
\end{figure}

The unit cell of the CLBM is shown on Fig. \ref{clbm} c) (labeled as $C$ structure). This cell was replicated (Fig. \ref{clbm} d) in order to describe the fibril structure where a number of collagen molecules are bundled together. The original water molecules was then removed and specific water clusters were inserted in the center of the structure.  The water cluster structures (H$_{2}$O)$_{n}$ ($1\leq n\leq8$)\footnote{It is possible the existence of more than a single symmetry for the same atomic set for $n\geq5$. Thus, we followed the choice of \cite{cybulski2007calculations} and considered only the structures (H$_{2}$O)$_{5-cyclic}$,(H$_{2}$O)$_{6-cage}$, (H$_{2}$O)$_{7-low}$, and (H$_{2}$O)$_{8-S4}$} were chosen based on the TIP4P model\cite{jorgensen1983comparison} from the Cambridge Cluster Database\cite{wales1998global}. Therefore we label the CLBM structures $C_{n}$ ($1\leq n\leq 8$). Two others variants of the $C$ structure were tested. The $C_{1}$ solvated with $10$ water molecules around the structure (named $C_{1s}$) and a more compact version of $C$ where only one water molecule was be fitted inside ($D$ labeled). These structures, $C$, $C_{1s}$, and $D$, represent the main sites where water clusters could be present in fibrils and collagen\cite{watercoll1,watercoll2,watercoll3}. The optimized unit cell lattice parameters are displayed on Table I. The structures displayed on Fig. \ref{c1sd0}. 

\begin{table}
\caption{Unit cell lattice parameters (\AA) for the CLBM structures. The symmetry was $P212121$ (orthorhombic).}
\begin{ruledtabular}
\begin{tabular}{cccc}
 & $C$ & $C_{1s}$ & $D_{1}$ \\ 
\hline 
$a=$ & $13.5$ & $15.0$ & $17.4$ \\ 
\hline 
$b=$ & $11.5$ & $12.5$ & $13.3$ \\ 
\hline 
$c=$ & $10.5$ & $20.0$ & $9.8$ \\ 
\end{tabular} 
\end{ruledtabular}
\end{table}

The $C$ models were optimized in a first step using molecular mechanics with the MMFF94s force field\cite{halgren1999mmff} implemented in the Avogadro software\cite{hanwell2012avogadro}. The D models were optimized in a first step using Hartree Fock with the $6\textrm{-}31\textrm{+}g(2d)$ basis implemented in the Gaussian software\cite{g03}. Then, Density Functional Theory (DFT)\cite{hohenberg1964inhomogeneous,kohn1965self} was used in order to obtain the equilibrium geometries and harmonic frequencies for the CLBM. The confinements were implemented in the CPMD program\cite{cpmd} using the BLYP functional\cite{lee1988development} augmented with dispersion corrections for the proper description of Van der Waals interactions\cite{von2005performance,lin2007library}. The cutoff energy was considered up to $100$ Ry in our simulations.

\begin{figure}[h!]
\includegraphics[width=8.0cm]{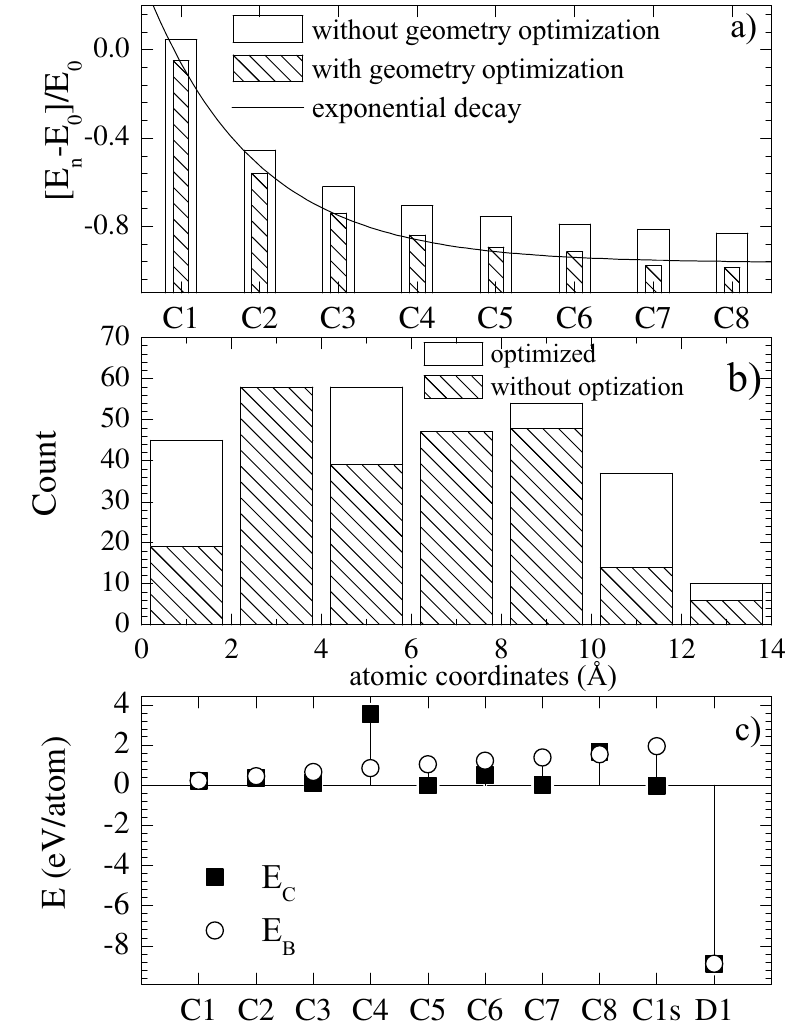}
\caption{a)Normalized total energy per atom with and without geometry optization with molecular mechanics for model $C$ with water clusters. b) Histogram for atomic coordinates of $C_{0}$ with and without geometry optization. c) Binding $E_{B}$ (circle) and couplig $E_{C}$ (square) symbols as defined in eqs. \ref{binding} and \ref{coupling}, respectively for the various structures considered. }\label{energies}
\end{figure}

The total energies for each structure were calculated. The data for the $C$ model are summarized on Fig. \ref{energies}a) where the energies were normalized to the value of $C_{0}$. We compared the energies with and without geometry optimization, since the non-optimized structures retained the original experimental structural parameters. However, the total energy decreased less than $20\%$ with optimization being the difference growing the greater the water cluster size was. It was observed that greater number of molecules per cluster lower the total energy of the cluster. In fact, we observed an exponential decay dependence on water content, $n$, according to

\begin{equation}\label{n}
\frac{E_{n}-E_{0}}{E_{0}}=-0.964+1.86e^{(-\frac{n}{1.388})}
\end{equation}
which is represented by a solid line in Fig.\ref{energies}a). It is important to notice that all cluster structures were stable ($E_{n}<E_{0}$). Park \textit{et al.}\cite{park3theoretical}  analyzed the structures and energies of the dimer to heptamer water clusters using the revised empirical potential function for conformational analysis, and realized that the cyclic structures of water clusters in general are more favorable than open structures, except for the heptamer. They also noted that the more hydrogen bonds a cluster has, the greater its stability. Thus, the stability of the chain of clusters increases by adding a water molecule to cluster. However, in our case it is important to notice that the energetic gain for $C_{4}<C<C_{8}$ is $\lesssim 8\%$. It could be stated that the energetic gain is minimum for $C>C_{4}$. Figure \ref{energies} b) shows a histogram of atomic distances for $C_{0}$ structure with and without geometry optimization in order to compare the effect of optimization in the experimentally determined original atomic positions. It was found that the shorter ($\lesssim 2$ \AA) and longer ($\gtrsim 10$ \AA) atomic relative distances presented the greatest variations. However it is important to notice that the original atomic connectivity and symmetry was unchanged.

To characterize the inter-water molecule interactions and the interactions between confined water molecules and collagen clusters we calculated the binding energy ($E_{B}$) and the coupling energy ($E_{C}$), respectively, according to ref.\cite{wang2009first}. $E_{B}$ is defined by

\begin{equation}\label{binding}
E_{B}=E_{CLBM}-(E_{0}+nE_{H_{2}O})
\end{equation}
and measures the average interaction between the free water molecules and their environment. $E_{CLBM}$, $E_{0}$, and $E_{H_{2}O}$ are the CLBM unit cell, $C_{0}$ or $D_{0}$, and water molecule energies of formation.

The interaction energy between amino acid cluster and water cluster within the CLBM unit cell could be estimated by $E_{C}$ as

\begin{equation}\label{coupling}
E_{C}=E_{CLBM}-(E_{0}+E_{nH_{2}O})
\end{equation}
where $E_{nH_{2}O}$ is the energy of the cluster composed by $n$ water molecules. Our definition differs from that of ref. \cite{wang2009first} by the minus sign and normalization to $n$.

Figure \ref{energies}c) presents $E_{B}$ for each structure. $E_{B}$ increased monotonically with $n$ for $C$ structure. Similar increasing behavior is observed in the literature. For example, Wang \textit{et al.}\cite{wang2009first} found a monotonically increasing energy as a number of water molecules for confined water in single-walled nanotubes. External work needs be done to introduce the water clusters inside the amino acid cage since $E_{B}>0$. Protein electrostatic forces and/or physiological electrochemical potential difference could furnish this energy. Thus, proteins locus with symmetry similar to $C$ structure are not viable places to catalyse the creation of water clusters.  Even so, these cages are feasible places for the insertion of water clusters. The solvating of the $C_{1}$ structure ($C_{1s}$) does not contribute to decreasing $E_{B}$. The $D$ structure containing one water molecule ($D_{1}$) presented the lowest $E_{B}$. One concludes that this compact structure is the viable one to trap $H_{2}O$ molecules.

The coupling energy $E_{C}$ for $C$ structures (Fig.\ref{energies}c) was found to be close to zero  independent of the water content, except for $n=4$ and $8$. It was found that the preferred cluster structure to pack water molecules was the $D_{1}$.  Tetrahedral $C_{4}$ and icosahedral $C_{8}$ water clusters appeared to have the greatest positive $E_{C}$. Thus, under normal conditions these clusters will be weakly interacting with the $C_{0}$ cage which implies in greater mobility. The coupling energy for the other $C$ structures are one a meV, falling into the energy scale of molecular vibrations. The $D_{1}$ structure was the more strongly coupled to the $D_{0}$ structure.

\begin{figure}[tbh!]
\includegraphics[width=8.0cm]{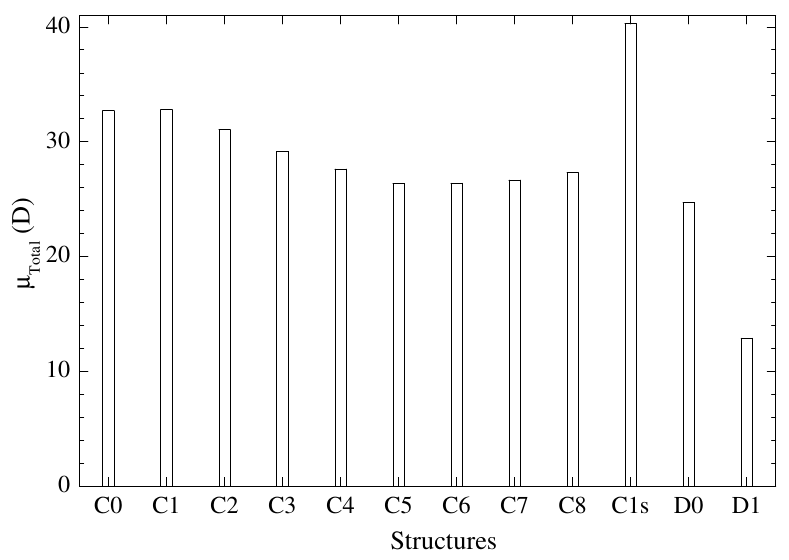}
\caption{Total dipole moment ($\mu_{Total}$) for each calculated structure.}\label{dipoles}
\end{figure}

The properties of water clusters are fundamental for understanding water in biological systems\cite{gelman2010structural}. Electrons transfer between biomolecular redox sites involves electron tunnelling through the intervening space, which can be facilitated in part by aqueous proton movement in the reverse direction. The rate of transfer depends on both the extent and the structure of the intervening space. In particular, proteins have much of their properties governed by interactions with water\cite{chaplin}. These, for example, establish the free-energy landscape that determines their folding, structure, stability, and activity\cite{halle2004protein}. Structured water clusters near redox cofactors accelerate electron transfer by producing strongly coupled tunnelling pathways between the electron donor and acceptor. More generally, water molecules can facilitate or disrupt this process through the degree of alignment of their dipoles and their capability to form water wires\cite{chaplin,chaplin2006we}.

The dipole moments calculated for each structure are shown on Fig.\ref{dipoles}. $C_{1s}$ displayed the greatest dipole moment. $D_{1}$ presented the lowest. The point charge $q$ and dipole moment, $\vec{\mu}$, average interaction potential is given by

\begin{equation}
U=-k^{2}\frac{q^2\mu^2}{3k_{B}Tr^{4}}
\end{equation}  
where $k$ is the Coulomb constant, $T$ is the temperature, and $r$ the distance among the point charge and $\vec{\mu}$. From Fig.\ref{dipoles} one concludes that a moving electron will be more strongly bound to the $C_{1s}$ structure. The $D_{1}$ structure will trap traveling electrons more weakly, favoring electron and small ion transport. From these qualitative clues it is expected that sites with $C$ symmetry will be preferred to host physiological processes were large water clusters and strongly bonded electrons are needed. The abundance of free tetrahedral water in the tumor tissue may be interpreted with our model framework considering that solid tumors have a tendency to enhance the permeability and retain large macromolecules ($>40$ kDa)\cite{tumor}. On the other hand, events where electron transport associated with lower hydration levels are the rule, will take place on loci with $D$ structure. These results confirm the hypothesis that confined water was altered in pathologic tissues, and due to some molecular arrangements of this water, it can influence the inflammatory response. 

Protein-water  interactions  are  known  to  play  a  critical  role in the function of several biomacromolecular systems including collagen tissues\cite{fathima2010structure}. Small  changes  in  structure  and  dynamical behavior of water molecules at the peptide-water interface can effectively change both the structure and dynamics of the protein function\cite{lima2012anharmonic}. Thus, it was realized that the presence of different water clusters plays an important role in collagen properties.

The properties of the structures proposed within the CLBM framework enabled  us to qualitatively correlate it to some important trends observed experimentally in inflammatory tissues. We found that a tetrahedral $C_{4}$ structure is weakly bound to the amino acid cage, being it the smallest stable structure with water cluster. The tetrahedral water inside this structure has high mobility. This correlates with literature findings reporting occurrence of tetrahedral water in pathological tissues. These clusters are the loci of, e.g.,  biochemical reactions, e.g., in the inflammatory cascade where strongly bounded electrons play very important role. We also found that $D$-like cages with one water molecule trap electrons more strongly being suitable ways for water and small ion transport. The CLBM is a very promising model for more complexes physical/biochemical properties of tissues. Water or other biofluids micro-perfusion, Young's modulus at microscale, electron/proton transport,  optical and electronic spectra, vibrational (Raman and IR)  spectra are some examples of   properties which could be simulated using this model. It is important to stress that more  studies need be conducted to correctly describe these others tissue properties. More experimental results will be explained and previsions about effects on tissues could be made.

\textbf{Acknowledgements.} The authors are grateful to the Brazilian agencies FAPESP, CAPES, and CNPq  for financial support and to the Multiuser Central Facilities of UFABC and CENAPAD for experimental and computational support, respectively.  E. Sato and H. Martinho are also grateful to Prof. Caetano Miranda (UFABC) for the critical discussions concerning the building of model.

\bibliography{water}

\begin{thebibliography}{44}%
\makeatletter
\providecommand \@ifxundefined [1]{%
 \@ifx{#1\undefined}
}%
\providecommand \@ifnum [1]{%
 \ifnum #1\expandafter \@firstoftwo
 \else \expandafter \@secondoftwo
 \fi
}%
\providecommand \@ifx [1]{%
 \ifx #1\expandafter \@firstoftwo
 \else \expandafter \@secondoftwo
 \fi
}%
\providecommand \natexlab [1]{#1}%
\providecommand \enquote  [1]{``#1''}%
\providecommand \bibnamefont  [1]{#1}%
\providecommand \bibfnamefont [1]{#1}%
\providecommand \citenamefont [1]{#1}%
\providecommand \href@noop [0]{\@secondoftwo}%
\providecommand \href [0]{\begingroup \@sanitize@url \@href}%
\providecommand \@href[1]{\@@startlink{#1}\@@href}%
\providecommand \@@href[1]{\endgroup#1\@@endlink}%
\providecommand \@sanitize@url [0]{\catcode `\\12\catcode `\$12\catcode
  `\&12\catcode `\#12\catcode `\^12\catcode `\_12\catcode `\%12\relax}%
\providecommand \@@startlink[1]{}%
\providecommand \@@endlink[0]{}%
\providecommand \url  [0]{\begingroup\@sanitize@url \@url }%
\providecommand \@url [1]{\endgroup\@href {#1}{\urlprefix }}%
\providecommand \urlprefix  [0]{URL }%
\providecommand \Eprint [0]{\href }%
\providecommand \doibase [0]{http://dx.doi.org/}%
\providecommand \selectlanguage [0]{\@gobble}%
\providecommand \bibinfo  [0]{\@secondoftwo}%
\providecommand \bibfield  [0]{\@secondoftwo}%
\providecommand \translation [1]{[#1]}%
\providecommand \BibitemOpen [0]{}%
\providecommand \bibitemStop [0]{}%
\providecommand \bibitemNoStop [0]{.\EOS\space}%
\providecommand \EOS [0]{\spacefactor3000\relax}%
\providecommand \BibitemShut  [1]{\csname bibitem#1\endcsname}%
\let\auto@bib@innerbib\@empty
\bibitem [{\citenamefont {Chaplin}(2011)}]{chaplin}%
  \BibitemOpen
  \bibfield  {author} {\bibinfo {author} {\bibfnamefont {M.}~\bibnamefont
  {Chaplin}},\ }\enquote {\bibinfo {title} {Water: The forgotten biological
  molecule},}\ \ (\bibinfo  {publisher} {Pan Stanford Publishing Pte. Ltd.},\
  \bibinfo {year} {2011})\ Chap.\ \bibinfo {chapter} {The structure of water
  molecule}\BibitemShut {NoStop}%
\bibitem [{\citenamefont {Leikin}\ \emph {et~al.}(1997)\citenamefont {Leikin},
  \citenamefont {Parsegian}, \citenamefont {Yang},\ and\ \citenamefont
  {Walrafen}}]{leikin1997raman}%
  \BibitemOpen
  \bibfield  {author} {\bibinfo {author} {\bibfnamefont {S.}~\bibnamefont
  {Leikin}}, \bibinfo {author} {\bibfnamefont {V.}~\bibnamefont {Parsegian}},
  \bibinfo {author} {\bibfnamefont {W.-H.}\ \bibnamefont {Yang}}, \ and\
  \bibinfo {author} {\bibfnamefont {G.}~\bibnamefont {Walrafen}},\ }\href@noop
  {} {\bibfield  {journal} {\bibinfo  {journal} {Proc. Nat. Acad. Sci.}\
  }\textbf {\bibinfo {volume} {94}},\ \bibinfo {pages} {11312} (\bibinfo {year}
  {1997})}\BibitemShut {NoStop}%
\bibitem [{\citenamefont {Carvalho}\ \emph {et~al.}(2011)\citenamefont
  {Carvalho}, \citenamefont {Sato}, \citenamefont {Almeida},\ and\
  \citenamefont {da~Martinho}}]{e2011diagnosis}%
  \BibitemOpen
  \bibfield  {author} {\bibinfo {author} {\bibfnamefont {L.~F. C.~S.}\
  \bibnamefont {Carvalho}}, \bibinfo {author} {\bibfnamefont {E.~T.}\
  \bibnamefont {Sato}}, \bibinfo {author} {\bibfnamefont {J.~D.}\ \bibnamefont
  {Almeida}}, \ and\ \bibinfo {author} {\bibfnamefont {H.~S.}\ \bibnamefont
  {da~Martinho}},\ }\href@noop {} {\bibfield  {journal} {\bibinfo  {journal}
  {Theor. Chem. Acc.}\ }\textbf {\bibinfo {volume} {130}},\ \bibinfo {pages}
  {1221} (\bibinfo {year} {2011})}\BibitemShut {NoStop}%
\bibitem [{\citenamefont {Movasaghi}\ \emph {et~al.}(2007)\citenamefont
  {Movasaghi}, \citenamefont {Rehman},\ and\ \citenamefont
  {Rehman}}]{movasaghi2007raman}%
  \BibitemOpen
  \bibfield  {author} {\bibinfo {author} {\bibfnamefont {Z.}~\bibnamefont
  {Movasaghi}}, \bibinfo {author} {\bibfnamefont {S.}~\bibnamefont {Rehman}}, \
  and\ \bibinfo {author} {\bibfnamefont {I.~U.}\ \bibnamefont {Rehman}},\
  }\href@noop {} {\bibfield  {journal} {\bibinfo  {journal} {Appl. Spectrosc.
  Rev.}\ }\textbf {\bibinfo {volume} {42}},\ \bibinfo {pages} {493} (\bibinfo
  {year} {2007})}\BibitemShut {NoStop}%
\bibitem [{\citenamefont {Cybulski}\ and\ \citenamefont
  {Sadlej}(2007)}]{cybulski2007calculations}%
  \BibitemOpen
  \bibfield  {author} {\bibinfo {author} {\bibfnamefont {H.}~\bibnamefont
  {Cybulski}}\ and\ \bibinfo {author} {\bibfnamefont {J.}~\bibnamefont
  {Sadlej}},\ }\href@noop {} {\bibfield  {journal} {\bibinfo  {journal} {Chem.
  Phys.}\ }\textbf {\bibinfo {volume} {342}},\ \bibinfo {pages} {163} (\bibinfo
  {year} {2007})}\BibitemShut {NoStop}%
\bibitem [{\citenamefont {McIntyre}(2006)}]{mcintyre2006cell}%
  \BibitemOpen
  \bibfield  {author} {\bibinfo {author} {\bibfnamefont {G.}~\bibnamefont
  {McIntyre}},\ }\href@noop {} {\bibfield  {journal} {\bibinfo  {journal} {Med.
  Hypotheses}\ }\textbf {\bibinfo {volume} {66}},\ \bibinfo {pages} {518}
  (\bibinfo {year} {2006})}\BibitemShut {NoStop}%
\bibitem [{\citenamefont {Chaplin}(2006)}]{chaplin2006we}%
  \BibitemOpen
  \bibfield  {author} {\bibinfo {author} {\bibfnamefont {M.}~\bibnamefont
  {Chaplin}},\ }\href@noop {} {\bibfield  {journal} {\bibinfo  {journal} {Nat.
  Rev. Mol. Cell Bio.}\ }\textbf {\bibinfo {volume} {7}},\ \bibinfo {pages}
  {861} (\bibinfo {year} {2006})}\BibitemShut {NoStop}%
\bibitem [{\citenamefont {Martinho}(2013)}]{martinholivro}%
  \BibitemOpen
  \bibfield  {author} {\bibinfo {author} {\bibfnamefont {H.}~\bibnamefont
  {Martinho}},\ }\enquote {\bibinfo {title} {Cancer – cares, treatments and
  preventions},}\ \ (\bibinfo  {publisher} {iConcept Press Ltd.},\ \bibinfo
  {year} {2013})\ Chap.\ \bibinfo {chapter} {Advances in Raman-based optical
  biopsy}\BibitemShut {NoStop}%
\bibitem [{\citenamefont {Gniadecka}\ \emph {et~al.}(1997)\citenamefont
  {Gniadecka}, \citenamefont {Wulf}, \citenamefont {Mortensen}, \citenamefont
  {Nielsen},\ and\ \citenamefont {Christensen}}]{gniadecka1997diagnosis}%
  \BibitemOpen
  \bibfield  {author} {\bibinfo {author} {\bibfnamefont {M.}~\bibnamefont
  {Gniadecka}}, \bibinfo {author} {\bibfnamefont {H.}~\bibnamefont {Wulf}},
  \bibinfo {author} {\bibfnamefont {N.~N.}\ \bibnamefont {Mortensen}}, \bibinfo
  {author} {\bibfnamefont {O.~F.}\ \bibnamefont {Nielsen}}, \ and\ \bibinfo
  {author} {\bibfnamefont {D.~H.}\ \bibnamefont {Christensen}},\ }\href@noop {}
  {\bibfield  {journal} {\bibinfo  {journal} {J. Raman Spectrosc.}\ }\textbf
  {\bibinfo {volume} {28}},\ \bibinfo {pages} {125} (\bibinfo {year}
  {1997})}\BibitemShut {NoStop}%
\bibitem [{\citenamefont {Xie}\ \emph {et~al.}(2012)\citenamefont {Xie},
  \citenamefont {Xie}, \citenamefont {Kinnings},\ and\ \citenamefont
  {Bourne}}]{comp_pharma}%
  \BibitemOpen
  \bibfield  {author} {\bibinfo {author} {\bibfnamefont {L.}~\bibnamefont
  {Xie}}, \bibinfo {author} {\bibfnamefont {L.}~\bibnamefont {Xie}}, \bibinfo
  {author} {\bibfnamefont {S.~L.}\ \bibnamefont {Kinnings}}, \ and\ \bibinfo
  {author} {\bibfnamefont {P.~E.}\ \bibnamefont {Bourne}},\ }\href@noop {}
  {\bibfield  {journal} {\bibinfo  {journal} {Annu. Rev. Pharmacol. Toxicol.}\
  }\textbf {\bibinfo {volume} {52}},\ \bibinfo {pages} {361} (\bibinfo {year}
  {2012})}\BibitemShut {NoStop}%
\bibitem [{\citenamefont {Steinhauser}\ and\ \citenamefont
  {Hiermaier}(2009)}]{comp_matsci}%
  \BibitemOpen
  \bibfield  {author} {\bibinfo {author} {\bibfnamefont {M.~O.}\ \bibnamefont
  {Steinhauser}}\ and\ \bibinfo {author} {\bibfnamefont {S.}~\bibnamefont
  {Hiermaier}},\ }\href@noop {} {\bibfield  {journal} {\bibinfo  {journal}
  {Int. J. Mol. Sci.}\ }\textbf {\bibinfo {volume} {10}},\ \bibinfo {pages}
  {5135} (\bibinfo {year} {2009})}\BibitemShut {NoStop}%
\bibitem [{\citenamefont {Yazid}\ \emph {et~al.}(2009)\citenamefont {Yazid},
  \citenamefont {Abdelkader},\ and\ \citenamefont {Abdelmadjid}}]{comp_mech}%
  \BibitemOpen
  \bibfield  {author} {\bibinfo {author} {\bibfnamefont {A.}~\bibnamefont
  {Yazid}}, \bibinfo {author} {\bibfnamefont {N.}~\bibnamefont {Abdelkader}}, \
  and\ \bibinfo {author} {\bibfnamefont {H.}~\bibnamefont {Abdelmadjid}},\
  }\href@noop {} {\bibfield  {journal} {\bibinfo  {journal} {Appl. Math.
  Model.}\ }\textbf {\bibinfo {volume} {33}},\ \bibinfo {pages} {4269 }
  (\bibinfo {year} {2009})}\BibitemShut {NoStop}%
\bibitem [{\citenamefont {L{\"u}}\ \emph {et~al.}(2013)\citenamefont {L{\"u}},
  \citenamefont {Liu}, \citenamefont {Gao}, \citenamefont {Huo}, \citenamefont
  {Kang}, \citenamefont {Chen}, \citenamefont {Sun}, \citenamefont {Chen},
  \citenamefont {Luo},\ and\ \citenamefont {Long}}]{lu2013asymmetric}%
  \BibitemOpen
  \bibfield  {author} {\bibinfo {author} {\bibfnamefont {D.}~\bibnamefont
  {L{\"u}}}, \bibinfo {author} {\bibfnamefont {X.}~\bibnamefont {Liu}},
  \bibinfo {author} {\bibfnamefont {Y.}~\bibnamefont {Gao}}, \bibinfo {author}
  {\bibfnamefont {B.}~\bibnamefont {Huo}}, \bibinfo {author} {\bibfnamefont
  {Y.}~\bibnamefont {Kang}}, \bibinfo {author} {\bibfnamefont {J.}~\bibnamefont
  {Chen}}, \bibinfo {author} {\bibfnamefont {S.}~\bibnamefont {Sun}}, \bibinfo
  {author} {\bibfnamefont {L.}~\bibnamefont {Chen}}, \bibinfo {author}
  {\bibfnamefont {X.}~\bibnamefont {Luo}}, \ and\ \bibinfo {author}
  {\bibfnamefont {M.}~\bibnamefont {Long}},\ }\href@noop {} {\bibfield
  {journal} {\bibinfo  {journal} {PloS one}\ }\textbf {\bibinfo {volume} {8}},\
  \bibinfo {pages} {e74563} (\bibinfo {year} {2013})}\BibitemShut {NoStop}%
\bibitem [{\citenamefont {Groves}\ \emph {et~al.}(2013)\citenamefont {Groves},
  \citenamefont {Coulman}, \citenamefont {Birchall},\ and\ \citenamefont
  {Evans}}]{Groves2013167}%
  \BibitemOpen
  \bibfield  {author} {\bibinfo {author} {\bibfnamefont {R.~B.}\ \bibnamefont
  {Groves}}, \bibinfo {author} {\bibfnamefont {S.~A.}\ \bibnamefont {Coulman}},
  \bibinfo {author} {\bibfnamefont {J.~C.}\ \bibnamefont {Birchall}}, \ and\
  \bibinfo {author} {\bibfnamefont {S.~L.}\ \bibnamefont {Evans}},\ }\href@noop
  {} {\bibfield  {journal} {\bibinfo  {journal} {J. Mech. Behav. Biomed.
  Mater.}\ }\textbf {\bibinfo {volume} {18}},\ \bibinfo {pages} {167 }
  (\bibinfo {year} {2013})}\BibitemShut {NoStop}%
\bibitem [{\citenamefont {Ren}\ \emph {et~al.}(2010)\citenamefont {Ren},
  \citenamefont {Liang}, \citenamefont {Qu}, \citenamefont {Li}, \citenamefont
  {Lu},\ and\ \citenamefont {Tian}}]{MC1}%
  \BibitemOpen
  \bibfield  {author} {\bibinfo {author} {\bibfnamefont {N.}~\bibnamefont
  {Ren}}, \bibinfo {author} {\bibfnamefont {J.}~\bibnamefont {Liang}}, \bibinfo
  {author} {\bibfnamefont {X.}~\bibnamefont {Qu}}, \bibinfo {author}
  {\bibfnamefont {J.}~\bibnamefont {Li}}, \bibinfo {author} {\bibfnamefont
  {B.}~\bibnamefont {Lu}}, \ and\ \bibinfo {author} {\bibfnamefont
  {J.}~\bibnamefont {Tian}},\ }\href {\doibase 10.1364/OE.18.006811} {\bibfield
   {journal} {\bibinfo  {journal} {Opt. Express}\ }\textbf {\bibinfo {volume}
  {18}},\ \bibinfo {pages} {6811} (\bibinfo {year} {2010})}\BibitemShut
  {NoStop}%
\bibitem [{\citenamefont {Zhu}\ and\ \citenamefont {Liu}(2013)}]{MC2}%
  \BibitemOpen
  \bibfield  {author} {\bibinfo {author} {\bibfnamefont {C.}~\bibnamefont
  {Zhu}}\ and\ \bibinfo {author} {\bibfnamefont {Q.}~\bibnamefont {Liu}},\
  }\href {\doibase 10.1117/1.JBO.18.5.050902} {\bibfield  {journal} {\bibinfo
  {journal} {J. Biom. Opt.}\ }\textbf {\bibinfo {volume} {18}},\ \bibinfo
  {pages} {050902} (\bibinfo {year} {2013})}\BibitemShut {NoStop}%
\bibitem [{\citenamefont {Mogilner}\ \emph {et~al.}(2002)\citenamefont
  {Mogilner}, \citenamefont {Ruderman},\ and\ \citenamefont
  {Grigera}}]{Mogilner2002209}%
  \BibitemOpen
  \bibfield  {author} {\bibinfo {author} {\bibfnamefont {I.~G.}\ \bibnamefont
  {Mogilner}}, \bibinfo {author} {\bibfnamefont {G.}~\bibnamefont {Ruderman}},
  \ and\ \bibinfo {author} {\bibfnamefont {J.}~\bibnamefont {Grigera}},\ }\href
  {\doibase http://dx.doi.org/10.1016/S1093-3263(02)00145-6} {\bibfield
  {journal} {\bibinfo  {journal} {J. Mol. Graphics Modell.}\ }\textbf {\bibinfo
  {volume} {21}},\ \bibinfo {pages} {209 } (\bibinfo {year}
  {2002})}\BibitemShut {NoStop}%
\bibitem [{\citenamefont {Simone}\ \emph {et~al.}(2008)\citenamefont {Simone},
  \citenamefont {Vitagliano},\ and\ \citenamefont {Berisio}}]{DeSimone2008121}%
  \BibitemOpen
  \bibfield  {author} {\bibinfo {author} {\bibfnamefont {A.~D.}\ \bibnamefont
  {Simone}}, \bibinfo {author} {\bibfnamefont {L.}~\bibnamefont {Vitagliano}},
  \ and\ \bibinfo {author} {\bibfnamefont {R.}~\bibnamefont {Berisio}},\ }\href
  {\doibase http://dx.doi.org/10.1016/j.bbrc.2008.04.190} {\bibfield  {journal}
  {\bibinfo  {journal} {Biochem. Bioph. Res. Co}\ }\textbf {\bibinfo {volume}
  {372}},\ \bibinfo {pages} {121 } (\bibinfo {year} {2008})}\BibitemShut
  {NoStop}%
\bibitem [{\citenamefont {Bodian}\ \emph {et~al.}(2011)\citenamefont {Bodian},
  \citenamefont {Radmer}, \citenamefont {Holbert},\ and\ \citenamefont
  {Klein}}]{terimd}%
  \BibitemOpen
  \bibfield  {author} {\bibinfo {author} {\bibfnamefont {D.~L.}\ \bibnamefont
  {Bodian}}, \bibinfo {author} {\bibfnamefont {R.~J.}\ \bibnamefont {Radmer}},
  \bibinfo {author} {\bibfnamefont {S.}~\bibnamefont {Holbert}}, \ and\
  \bibinfo {author} {\bibfnamefont {T.~E.}\ \bibnamefont {Klein}},\ }in\
  \href@noop {} {\emph {\bibinfo {booktitle} {Pacific Symposium on
  Biocomputing}}}\ (\bibinfo {organization} {World Scientific},\ \bibinfo
  {year} {2011})\ pp.\ \bibinfo {pages} {193--204}\BibitemShut {NoStop}%
\bibitem [{\citenamefont {Streeter}\ and\ \citenamefont
  {de~Leeuw}(2011)}]{C0SM01192D}%
  \BibitemOpen
  \bibfield  {author} {\bibinfo {author} {\bibfnamefont {I.}~\bibnamefont
  {Streeter}}\ and\ \bibinfo {author} {\bibfnamefont {N.~H.}\ \bibnamefont
  {de~Leeuw}},\ }\href {\doibase 10.1039/C0SM01192D} {\bibfield  {journal}
  {\bibinfo  {journal} {Soft Matter}\ }\textbf {\bibinfo {volume} {7}},\
  \bibinfo {pages} {3373} (\bibinfo {year} {2011})}\BibitemShut {NoStop}%
\bibitem [{\citenamefont {Suárez}\ \emph {et~al.}(2009)\citenamefont
  {Suárez}, \citenamefont {Díaz},\ and\ \citenamefont
  {Suárez}}]{suarezqmmm}%
  \BibitemOpen
  \bibfield  {author} {\bibinfo {author} {\bibfnamefont {E.}~\bibnamefont
  {Suárez}}, \bibinfo {author} {\bibfnamefont {N.}~\bibnamefont {Díaz}}, \
  and\ \bibinfo {author} {\bibfnamefont {D.}~\bibnamefont {Suárez}},\
  }\href@noop {} {\bibfield  {journal} {\bibinfo  {journal} {J. Chem. Theory
  Comput.}\ }\textbf {\bibinfo {volume} {5}},\ \bibinfo {pages} {1667}
  (\bibinfo {year} {2009})}\BibitemShut {NoStop}%
\bibitem [{\citenamefont {Bella}\ \emph
  {et~al.}(1995{\natexlab{a}})\citenamefont {Bella}, \citenamefont {Brodsky},\
  and\ \citenamefont {Berman}}]{bella1995hydration}%
  \BibitemOpen
  \bibfield  {author} {\bibinfo {author} {\bibfnamefont {J.}~\bibnamefont
  {Bella}}, \bibinfo {author} {\bibfnamefont {B.}~\bibnamefont {Brodsky}}, \
  and\ \bibinfo {author} {\bibfnamefont {H.~M.}\ \bibnamefont {Berman}},\
  }\href@noop {} {\bibfield  {journal} {\bibinfo  {journal} {Structure}\
  }\textbf {\bibinfo {volume} {3}},\ \bibinfo {pages} {893} (\bibinfo {year}
  {1995}{\natexlab{a}})}\BibitemShut {NoStop}%
\bibitem [{Note1()}]{Note1}%
  \BibitemOpen
  \bibinfo {note} {It is possible the existence of more than a single symmetry
  for the same atomic set for $n\geq 5$. Thus, we followed the choice of \cite
  {cybulski2007calculations} and considered only the structures
  (H$_{2}$O)$_{5-cyclic}$,(H$_{2}$O)$_{6-cage}$, (H$_{2}$O)$_{7-low}$, and
  (H$_{2}$O)$_{8-S4}$}\BibitemShut {NoStop}%
\bibitem [{\citenamefont {Jorgensen}\ \emph {et~al.}(1983)\citenamefont
  {Jorgensen}, \citenamefont {Chandrasekhar}, \citenamefont {Madura},
  \citenamefont {Impey},\ and\ \citenamefont
  {Klein}}]{jorgensen1983comparison}%
  \BibitemOpen
  \bibfield  {author} {\bibinfo {author} {\bibfnamefont {W.~L.}\ \bibnamefont
  {Jorgensen}}, \bibinfo {author} {\bibfnamefont {J.}~\bibnamefont
  {Chandrasekhar}}, \bibinfo {author} {\bibfnamefont {J.~D.}\ \bibnamefont
  {Madura}}, \bibinfo {author} {\bibfnamefont {R.~W.}\ \bibnamefont {Impey}}, \
  and\ \bibinfo {author} {\bibfnamefont {M.~L.}\ \bibnamefont {Klein}},\
  }\href@noop {} {\bibfield  {journal} {\bibinfo  {journal} {J. Chem. Phys.}\
  }\textbf {\bibinfo {volume} {79}},\ \bibinfo {pages} {926} (\bibinfo {year}
  {1983})}\BibitemShut {NoStop}%
\bibitem [{\citenamefont {Wales}\ and\ \citenamefont
  {Hodges}(1998)}]{wales1998global}%
  \BibitemOpen
  \bibfield  {author} {\bibinfo {author} {\bibfnamefont {D.~J.}\ \bibnamefont
  {Wales}}\ and\ \bibinfo {author} {\bibfnamefont {M.~P.}\ \bibnamefont
  {Hodges}},\ }\href@noop {} {\bibfield  {journal} {\bibinfo  {journal} {Chem.
  Phys. Lett.}\ }\textbf {\bibinfo {volume} {286}},\ \bibinfo {pages} {65}
  (\bibinfo {year} {1998})}\BibitemShut {NoStop}%
\bibitem [{\citenamefont {Bella}\ \emph
  {et~al.}(1995{\natexlab{b}})\citenamefont {Bella}, \citenamefont {Brodsky},\
  and\ \citenamefont {Berman}}]{watercoll1}%
  \BibitemOpen
  \bibfield  {author} {\bibinfo {author} {\bibfnamefont {J.}~\bibnamefont
  {Bella}}, \bibinfo {author} {\bibfnamefont {B.}~\bibnamefont {Brodsky}}, \
  and\ \bibinfo {author} {\bibfnamefont {H.~M.}\ \bibnamefont {Berman}},\
  }\href {\doibase http://dx.doi.org/10.1016/S0969-2126(01)00224-6} {\bibfield
  {journal} {\bibinfo  {journal} {Structure}\ }\textbf {\bibinfo {volume}
  {3}},\ \bibinfo {pages} {893 } (\bibinfo {year}
  {1995}{\natexlab{b}})}\BibitemShut {NoStop}%
\bibitem [{\citenamefont {VanderSchee}\ and\ \citenamefont
  {Ooms}(2014)}]{watercoll2}%
  \BibitemOpen
  \bibfield  {author} {\bibinfo {author} {\bibfnamefont {C.~R.}\ \bibnamefont
  {VanderSchee}}\ and\ \bibinfo {author} {\bibfnamefont {K.~J.}\ \bibnamefont
  {Ooms}},\ }\href@noop {} {\bibfield  {journal} {\bibinfo  {journal} {J. Phys.
  Chem. B}\ } (\bibinfo {year} {2014})}\BibitemShut {NoStop}%
\bibitem [{\citenamefont {Kavukcuoglu}\ \emph {et~al.}(2012)\citenamefont
  {Kavukcuoglu}, \citenamefont {Li}, \citenamefont {Pleshko},\ and\
  \citenamefont {Uitto}}]{watercoll3}%
  \BibitemOpen
  \bibfield  {author} {\bibinfo {author} {\bibfnamefont {N.~B.}\ \bibnamefont
  {Kavukcuoglu}}, \bibinfo {author} {\bibfnamefont {Q.}~\bibnamefont {Li}},
  \bibinfo {author} {\bibfnamefont {N.}~\bibnamefont {Pleshko}}, \ and\
  \bibinfo {author} {\bibfnamefont {J.}~\bibnamefont {Uitto}},\ }\href
  {\doibase http://dx.doi.org/10.1016/j.matbio.2012.02.004} {\bibfield
  {journal} {\bibinfo  {journal} {Matrix Biol.}\ }\textbf {\bibinfo {volume}
  {31}},\ \bibinfo {pages} {246 } (\bibinfo {year} {2012})}\BibitemShut
  {NoStop}%
\bibitem [{\citenamefont {Halgren}(1999)}]{halgren1999mmff}%
  \BibitemOpen
  \bibfield  {author} {\bibinfo {author} {\bibfnamefont {T.~A.}\ \bibnamefont
  {Halgren}},\ }\href@noop {} {\bibfield  {journal} {\bibinfo  {journal} {J.
  Comput. Chem.}\ }\textbf {\bibinfo {volume} {20}},\ \bibinfo {pages} {720}
  (\bibinfo {year} {1999})}\BibitemShut {NoStop}%
\bibitem [{\citenamefont {Hanwell}\ \emph {et~al.}(2012)\citenamefont
  {Hanwell}, \citenamefont {Curtis}, \citenamefont {Lonie}, \citenamefont
  {Vandermeersch}, \citenamefont {Zurek},\ and\ \citenamefont
  {Hutchison}}]{hanwell2012avogadro}%
  \BibitemOpen
  \bibfield  {author} {\bibinfo {author} {\bibfnamefont {M.~D.}\ \bibnamefont
  {Hanwell}}, \bibinfo {author} {\bibfnamefont {D.~E.}\ \bibnamefont {Curtis}},
  \bibinfo {author} {\bibfnamefont {D.~C.}\ \bibnamefont {Lonie}}, \bibinfo
  {author} {\bibfnamefont {T.}~\bibnamefont {Vandermeersch}}, \bibinfo {author}
  {\bibfnamefont {E.}~\bibnamefont {Zurek}}, \ and\ \bibinfo {author}
  {\bibfnamefont {G.~R.}\ \bibnamefont {Hutchison}},\ }\href@noop {} {\bibfield
   {journal} {\bibinfo  {journal} {J. Cheminfor.}\ }\textbf {\bibinfo {volume}
  {4}},\ \bibinfo {pages} {1} (\bibinfo {year} {2012})}\BibitemShut {NoStop}%
\bibitem [{\citenamefont {\textit{et al.}}(2004)}]{g03}%
  \BibitemOpen
  \bibfield  {author} {\bibinfo {author} {\bibfnamefont {M.~J.~F.}\
  \bibnamefont {\textit{et al.}}},\ }\href@noop {} {\enquote {\bibinfo {title}
  {Gaussian 03, \uppercase{R}evision \uppercase{C}.02},}\ } (\bibinfo {year}
  {2004}),\ \bibinfo {note} {\uppercase{G}aussian, Inc., Wallingford, CT,
  2004}\BibitemShut {NoStop}%
\bibitem [{\citenamefont {Hohenberg}\ and\ \citenamefont
  {Kohn}(1964)}]{hohenberg1964inhomogeneous}%
  \BibitemOpen
  \bibfield  {author} {\bibinfo {author} {\bibfnamefont {P.}~\bibnamefont
  {Hohenberg}}\ and\ \bibinfo {author} {\bibfnamefont {W.}~\bibnamefont
  {Kohn}},\ }\href@noop {} {\bibfield  {journal} {\bibinfo  {journal} {Phys.
  Rev.}\ }\textbf {\bibinfo {volume} {136}},\ \bibinfo {pages} {B864} (\bibinfo
  {year} {1964})}\BibitemShut {NoStop}%
\bibitem [{\citenamefont {Kohn}\ and\ \citenamefont
  {Sham}(1965)}]{kohn1965self}%
  \BibitemOpen
  \bibfield  {author} {\bibinfo {author} {\bibfnamefont {W.}~\bibnamefont
  {Kohn}}\ and\ \bibinfo {author} {\bibfnamefont {L.~J.}\ \bibnamefont
  {Sham}},\ }\href@noop {} {\bibfield  {journal} {\bibinfo  {journal} {Phys.
  Rev.}\ }\textbf {\bibinfo {volume} {140}},\ \bibinfo {pages} {A1133}
  (\bibinfo {year} {1965})}\BibitemShut {NoStop}%
\bibitem [{cpm(2001)}]{cpmd}%
  \BibitemOpen
  \href@noop {} {\enquote {\bibinfo {title} {Cpmd,
  http://www.cpmd.org/,copyright ibm corp 1990-2008, copyright mpi für
  festkörperforschung stuttgart.}}\ } (\bibinfo {year}
  {1997-2001})\BibitemShut {NoStop}%
\bibitem [{\citenamefont {Lee}\ \emph {et~al.}(1988)\citenamefont {Lee},
  \citenamefont {Yang},\ and\ \citenamefont {Parr}}]{lee1988development}%
  \BibitemOpen
  \bibfield  {author} {\bibinfo {author} {\bibfnamefont {C.}~\bibnamefont
  {Lee}}, \bibinfo {author} {\bibfnamefont {W.}~\bibnamefont {Yang}}, \ and\
  \bibinfo {author} {\bibfnamefont {R.~G.}\ \bibnamefont {Parr}},\ }\href@noop
  {} {\bibfield  {journal} {\bibinfo  {journal} {Phys. Rev. B}\ }\textbf
  {\bibinfo {volume} {37}},\ \bibinfo {pages} {785} (\bibinfo {year}
  {1988})}\BibitemShut {NoStop}%
\bibitem [{\citenamefont {von Lilienfeld}\ \emph {et~al.}(2005)\citenamefont
  {von Lilienfeld}, \citenamefont {Tavernelli}, \citenamefont {Rothlisberger},\
  and\ \citenamefont {Sebastiani}}]{von2005performance}%
  \BibitemOpen
  \bibfield  {author} {\bibinfo {author} {\bibfnamefont {O.~A.}\ \bibnamefont
  {von Lilienfeld}}, \bibinfo {author} {\bibfnamefont {I.}~\bibnamefont
  {Tavernelli}}, \bibinfo {author} {\bibfnamefont {U.}~\bibnamefont
  {Rothlisberger}}, \ and\ \bibinfo {author} {\bibfnamefont {D.}~\bibnamefont
  {Sebastiani}},\ }\href@noop {} {\bibfield  {journal} {\bibinfo  {journal}
  {Phys. Rev. B}\ }\textbf {\bibinfo {volume} {71}},\ \bibinfo {pages} {195119}
  (\bibinfo {year} {2005})}\BibitemShut {NoStop}%
\bibitem [{\citenamefont {Lin}\ \emph {et~al.}(2007)\citenamefont {Lin},
  \citenamefont {Coutinho-Neto}, \citenamefont {Felsenheimer}, \citenamefont
  {von Lilienfeld}, \citenamefont {Tavernelli},\ and\ \citenamefont
  {Rothlisberger}}]{lin2007library}%
  \BibitemOpen
  \bibfield  {author} {\bibinfo {author} {\bibfnamefont {I.-C.}\ \bibnamefont
  {Lin}}, \bibinfo {author} {\bibfnamefont {M.~D.}\ \bibnamefont
  {Coutinho-Neto}}, \bibinfo {author} {\bibfnamefont {C.}~\bibnamefont
  {Felsenheimer}}, \bibinfo {author} {\bibfnamefont {O.~A.}\ \bibnamefont {von
  Lilienfeld}}, \bibinfo {author} {\bibfnamefont {I.}~\bibnamefont
  {Tavernelli}}, \ and\ \bibinfo {author} {\bibfnamefont {U.}~\bibnamefont
  {Rothlisberger}},\ }\href@noop {} {\bibfield  {journal} {\bibinfo  {journal}
  {Phys. Rev. B}\ }\textbf {\bibinfo {volume} {75}},\ \bibinfo {pages} {205131}
  (\bibinfo {year} {2007})}\BibitemShut {NoStop}%
\bibitem [{\citenamefont {Park}\ \emph {et~al.}(1982)\citenamefont {Park},
  \citenamefont {Kang}, \citenamefont {Yoon},\ and\ \citenamefont
  {Jhon}}]{park3theoretical}%
  \BibitemOpen
  \bibfield  {author} {\bibinfo {author} {\bibfnamefont {Y.~J.}\ \bibnamefont
  {Park}}, \bibinfo {author} {\bibfnamefont {Y.~K.}\ \bibnamefont {Kang}},
  \bibinfo {author} {\bibfnamefont {B.~J.}\ \bibnamefont {Yoon}}, \ and\
  \bibinfo {author} {\bibfnamefont {M.~S.}\ \bibnamefont {Jhon}},\ }\href@noop
  {} {\bibfield  {journal} {\bibinfo  {journal} {B. Korean Chem. Soc.}\
  }\textbf {\bibinfo {volume} {3}},\ \bibinfo {pages} {50} (\bibinfo {year}
  {1982})}\BibitemShut {NoStop}%
\bibitem [{\citenamefont {Wang}\ \emph {et~al.}(2009)\citenamefont {Wang},
  \citenamefont {Zhao}, \citenamefont {Li}, \citenamefont {Fang},\ and\
  \citenamefont {Lu}}]{wang2009first}%
  \BibitemOpen
  \bibfield  {author} {\bibinfo {author} {\bibfnamefont {L.}~\bibnamefont
  {Wang}}, \bibinfo {author} {\bibfnamefont {J.}~\bibnamefont {Zhao}}, \bibinfo
  {author} {\bibfnamefont {F.}~\bibnamefont {Li}}, \bibinfo {author}
  {\bibfnamefont {H.}~\bibnamefont {Fang}}, \ and\ \bibinfo {author}
  {\bibfnamefont {J.~P.}\ \bibnamefont {Lu}},\ }\href@noop {} {\bibfield
  {journal} {\bibinfo  {journal} {J. Phys. Chem. C}\ }\textbf {\bibinfo
  {volume} {113}},\ \bibinfo {pages} {5368} (\bibinfo {year}
  {2009})}\BibitemShut {NoStop}%
\bibitem [{\citenamefont {Gelman-Constantin}\ \emph {et~al.}(2010)\citenamefont
  {Gelman-Constantin}, \citenamefont {Carignano}, \citenamefont {Szleifer},
  \citenamefont {Marceca},\ and\ \citenamefont {Corti}}]{gelman2010structural}%
  \BibitemOpen
  \bibfield  {author} {\bibinfo {author} {\bibfnamefont {J.}~\bibnamefont
  {Gelman-Constantin}}, \bibinfo {author} {\bibfnamefont {M.~A.}\ \bibnamefont
  {Carignano}}, \bibinfo {author} {\bibfnamefont {I.}~\bibnamefont {Szleifer}},
  \bibinfo {author} {\bibfnamefont {E.~J.}\ \bibnamefont {Marceca}}, \ and\
  \bibinfo {author} {\bibfnamefont {H.~R.}\ \bibnamefont {Corti}},\ }\href@noop
  {} {\bibfield  {journal} {\bibinfo  {journal} {J. Chem. Phys.}\ }\textbf
  {\bibinfo {volume} {133}},\ \bibinfo {pages} {024506} (\bibinfo {year}
  {2010})}\BibitemShut {NoStop}%
\bibitem [{\citenamefont {Halle}(2004)}]{halle2004protein}%
  \BibitemOpen
  \bibfield  {author} {\bibinfo {author} {\bibfnamefont {B.}~\bibnamefont
  {Halle}},\ }\href@noop {} {\bibfield  {journal} {\bibinfo  {journal} {Philos.
  T. R. Soc. London. B}\ }\textbf {\bibinfo {volume} {359}},\ \bibinfo {pages}
  {1207} (\bibinfo {year} {2004})}\BibitemShut {NoStop}%
\bibitem [{\citenamefont {Fang}\ \emph {et~al.}(2012)\citenamefont {Fang},
  \citenamefont {Qin}, \citenamefont {Nakamura}, \citenamefont {Tsukigawa},
  \citenamefont {Shin},\ and\ \citenamefont {Maeda}}]{tumor}%
  \BibitemOpen
  \bibfield  {author} {\bibinfo {author} {\bibfnamefont {J.}~\bibnamefont
  {Fang}}, \bibinfo {author} {\bibfnamefont {H.}~\bibnamefont {Qin}}, \bibinfo
  {author} {\bibfnamefont {H.}~\bibnamefont {Nakamura}}, \bibinfo {author}
  {\bibfnamefont {K.}~\bibnamefont {Tsukigawa}}, \bibinfo {author}
  {\bibfnamefont {T.}~\bibnamefont {Shin}}, \ and\ \bibinfo {author}
  {\bibfnamefont {H.}~\bibnamefont {Maeda}},\ }\href {\doibase
  10.1111/j.1349-7006.2011.02178.x} {\bibfield  {journal} {\bibinfo  {journal}
  {Cancer Science}\ }\textbf {\bibinfo {volume} {103}},\ \bibinfo {pages} {535}
  (\bibinfo {year} {2012})}\BibitemShut {NoStop}%
\bibitem [{\citenamefont {Fathima}\ \emph {et~al.}(2010)\citenamefont
  {Fathima}, \citenamefont {Baias}, \citenamefont {Blumich},\ and\
  \citenamefont {Ramasami}}]{fathima2010structure}%
  \BibitemOpen
  \bibfield  {author} {\bibinfo {author} {\bibfnamefont {N.~N.}\ \bibnamefont
  {Fathima}}, \bibinfo {author} {\bibfnamefont {M.}~\bibnamefont {Baias}},
  \bibinfo {author} {\bibfnamefont {B.}~\bibnamefont {Blumich}}, \ and\
  \bibinfo {author} {\bibfnamefont {T.}~\bibnamefont {Ramasami}},\ }\href@noop
  {} {\bibfield  {journal} {\bibinfo  {journal} {Int. J. Biol. Macromol.}\
  }\textbf {\bibinfo {volume} {47}},\ \bibinfo {pages} {590} (\bibinfo {year}
  {2010})}\BibitemShut {NoStop}%
\bibitem [{\citenamefont {Lima}\ \emph {et~al.}(2012)\citenamefont {Lima},
  \citenamefont {Sato}, \citenamefont {Martins}, \citenamefont {Homem-de
  Mello}, \citenamefont {Lago}, \citenamefont {Coutinho-Neto}, \citenamefont
  {Ferreira}, \citenamefont {Giles}, \citenamefont {Pires},\ and\ \citenamefont
  {Martinho}}]{lima2012anharmonic}%
  \BibitemOpen
  \bibfield  {author} {\bibinfo {author} {\bibfnamefont {T.}~\bibnamefont
  {Lima}}, \bibinfo {author} {\bibfnamefont {E.}~\bibnamefont {Sato}}, \bibinfo
  {author} {\bibfnamefont {E.}~\bibnamefont {Martins}}, \bibinfo {author}
  {\bibfnamefont {P.}~\bibnamefont {Homem-de Mello}}, \bibinfo {author}
  {\bibfnamefont {A.}~\bibnamefont {Lago}}, \bibinfo {author} {\bibfnamefont
  {M.}~\bibnamefont {Coutinho-Neto}}, \bibinfo {author} {\bibfnamefont
  {F.}~\bibnamefont {Ferreira}}, \bibinfo {author} {\bibfnamefont
  {C.}~\bibnamefont {Giles}}, \bibinfo {author} {\bibfnamefont
  {M.}~\bibnamefont {Pires}}, \ and\ \bibinfo {author} {\bibfnamefont
  {H.}~\bibnamefont {Martinho}},\ }\href@noop {} {\bibfield  {journal}
  {\bibinfo  {journal} {J. Phys.-Condens. Mat.}\ }\textbf {\bibinfo {volume}
  {24}},\ \bibinfo {pages} {195104} (\bibinfo {year} {2012})}\BibitemShut
  {NoStop}%
\end{thebibliography}%

\end{document}